\documentclass[final,5p,times,twocolumn]{elsarticle}
\usepackage{upgreek}

\usepackage{amssymb}
\usepackage{amsmath}

\usepackage{booktabs}
\usepackage{hyperref}

\journal{Nuclear Insruments and Methods A}

\begin{document}

\begin{frontmatter}



\title{Radiation damage and recovery of SiPMs exposed to low-momentum muon beams for innovative scintillation-based muon beam monitors}




\author[unipv,infn]{R. Rossini\corref{cor1}}
\ead{riccardo.rossini@infn.it}
\cortext[cor1]{Corresponding author}

\author[isis]{S. Biswas}
\author[isis]{A.D. Hillier}
\author[isis]{J.S. Lord}
\author[unipv,infn]{L. Malliolio}
\author[unipv,infn]{A. Menegolli}
\author[unipv]{W. Mourad}
\author[isis]{D.E. Pooley}
\author[infn]{M.C. Prata}
\author[isis]{L. Quintieri}
\author[infn]{M. Rossella}
\author[unipv]{F. Zamperoni}

\affiliation[unipv]{
    organization={Department of Physics "A. Volta", University of Pavia},
    city={Pavia},
    country={Italy}
}

\affiliation[infn]{
    organization={INFN sezione di Pavia},
    city={Pavia},
    country={Italy}
}

\affiliation[isis]{
    organization={ISIS Neutron and Muon Source, Science and Technology Facilities Council},
    city={Didcot},
    country={United Kingdom}
}

\begin{abstract}
Silicon PhotoMultipliers (SiPMs) are among the standard devices for light detection in scintillator detectors. Some detector designs require these devices to be placed in radioactive areas or directly exposed to particle beams. Specifically, the damage from exposure to low-momentum muons is particularly important, as they have a high energy deposit and they have several applications in material science and particle physics. Eight SiPM models from different manufacturers were exposed to 29 MeV/c positive muons, with an intensity of $\mathcal{O}(10^6)$ muons/s for 13 hours, maximising the charge deposited in the depletion region. I-V curves were measured before/after irradiation and annealing; DCR and crosstalk analyses are ongoing, to extract dark count rate and optical crosstalk. Despite showing a measurable increase in dark current, the devices preserved operational stability. Thermal annealing has also been tested by heating the devices at 125°C, showing a recovery of the optimal SiPM features. This work is part of a collaboration between the University of Pavia and ISIS, to develop detectors for low-momentum muons (< 40 MeV/c), based on thin scintillators.
\end{abstract}




\end{frontmatter}


\section{Introduction}
    \begin{figure}
        \centering
        \includegraphics[width=\linewidth]{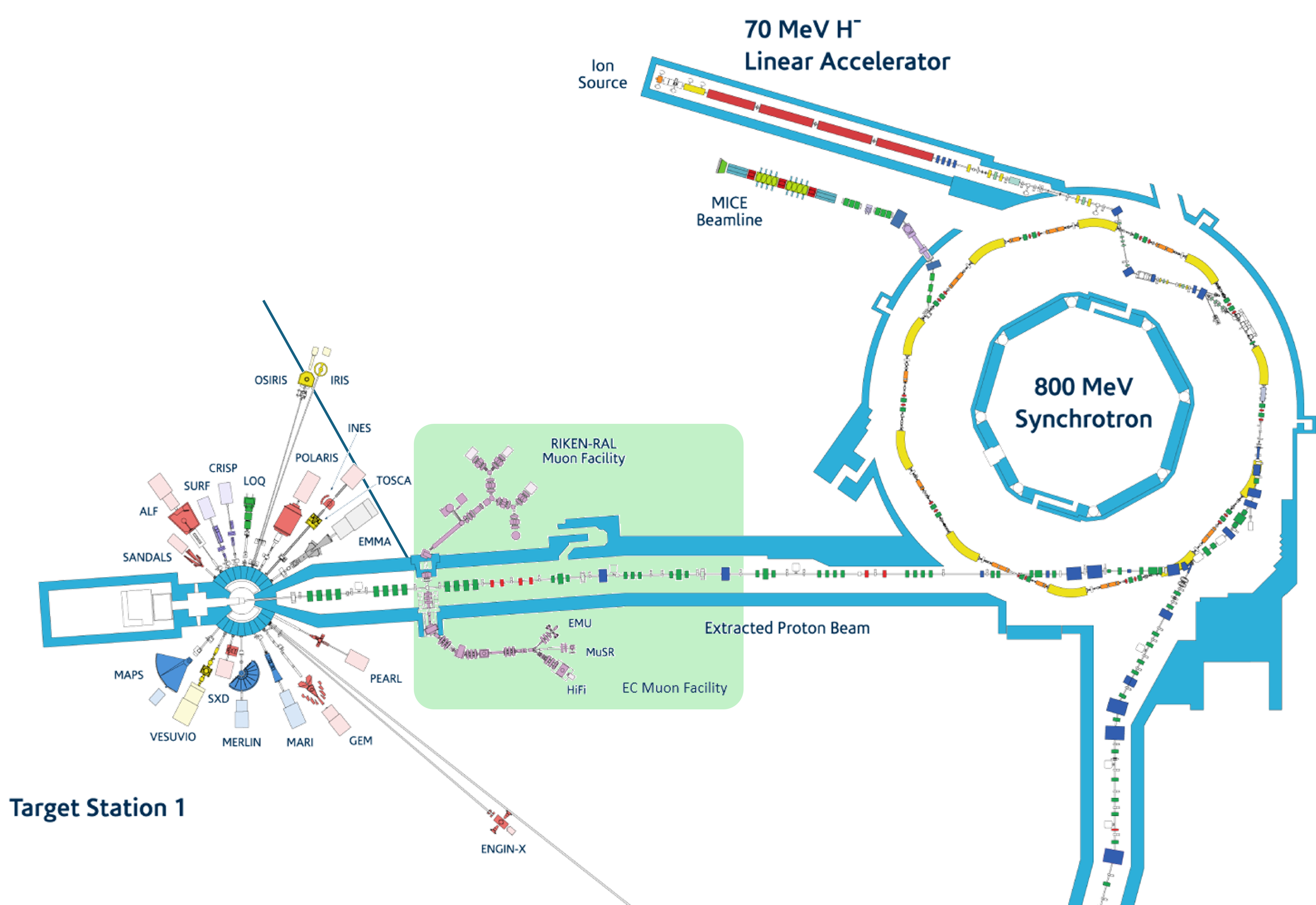}
        \caption{Close-up scheme of the ISIS synchrotron and Target Station 1. The ISIS Muon Facilties (RIKEN-Ral Ports 1-4 and EC beamlines EMU, HiFI and MuSR) are highlighted in green. Adapted from \cite{RIKEN}}.
        \label{fig:isis}
    \end{figure}
    
\begin{figure}
        \centering
        \includegraphics[width=\linewidth]{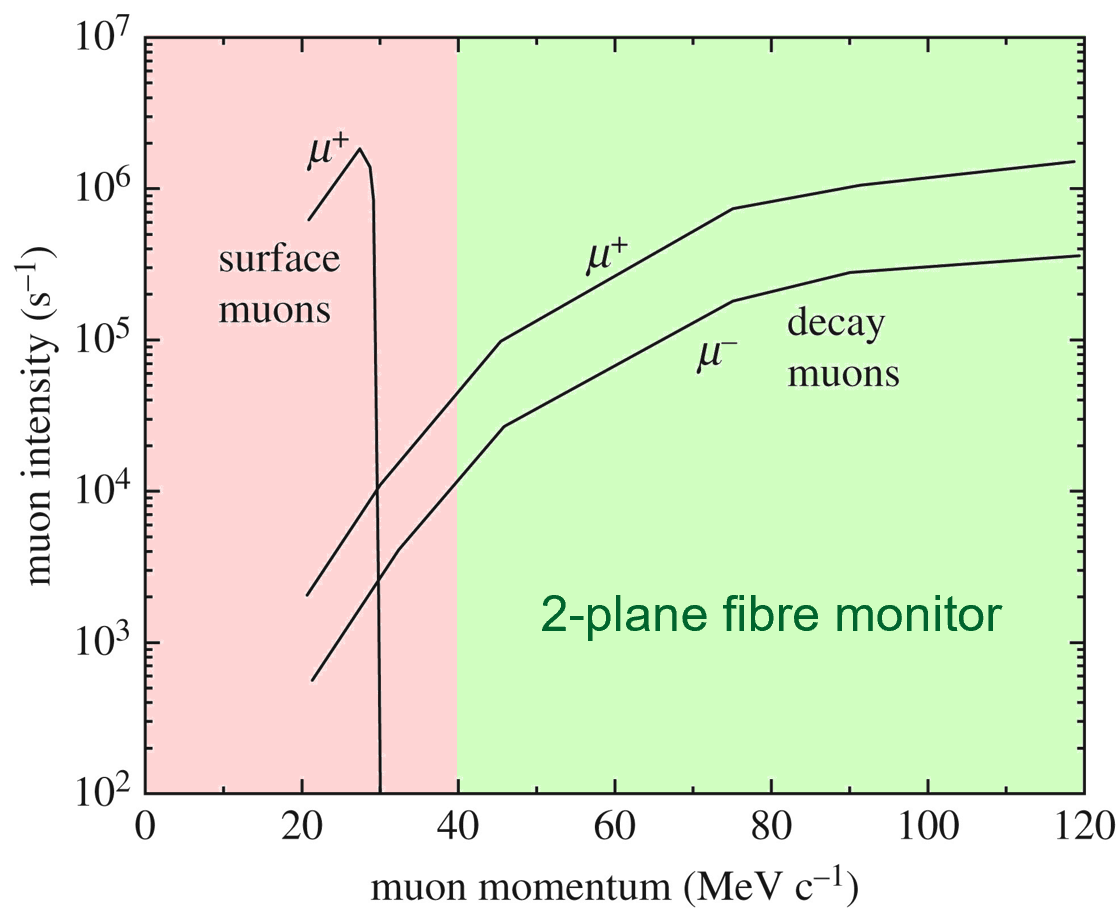}  
        \caption{Estimated intensity of surface $\mu^+$ and decay $\mu^{+/-}$ delivered by the RIKEN-RAL facility. The beam monitors developed by the FAMU and CHNet-MAXI collaborations, with a dense 2-fibre configuration, can only be used over $\sim 40$ MeV/c, whereas a new design has to be adopted for lower momentum muons. Adapted from \cite{RIKEN}.}
        \label{fig:int}
    \end{figure}

    The ISIS Neutron and Muon Source features two different muon facilities\cite{RIKEN}, with a total of 7 beamlines, as shown in Figure \ref{fig:isis}. The EC muon facility is capable of delivering surface $\mu^+$ beams, coming from the decay of $\pi^+$ at rest in the production target, with high intensity and fixed momentum around 29 MeV/c. On the other hand, RIKEN-RAL is a decay muon facility: in addition to surface muons, it features a decay solenoid, which transports $\pi^{+/-}$ and allows them to decay in flight. For this reason, muons of both polarities and in momentum range 17-120 MeV/c can be delivered by RIKEN-RAL. Figure \ref{fig:int} shows the values of beam intensity and momenta available at RIKEN-RAL.
    
    Several groups have developed beam monitors for muon beamlines. The most used design for low-momentum beams is a set of two planes of square scintillating fibres, either in a dense plane geometry\cite{Hodo1,Hodo2,hodomaxi}, or with gaps between fibres\cite{dalmaso:2023, papa:2015, papa:2019, rossini2024hodo}. The first geometry allows for the highest spatial resolution, but it can be more difficult to assemble and it can interfere with the beam or block it completely; the second one is less invasive, but it needs thorough calibration to extract useful information, and in general it has worse spatial resolution. 
    
    A new geometry for scintillator-based beam monitors is being designed and tested as part of a collaboration between Pavia research institutions (University, INFN and CNAO) and ISIS. It basically involves a single thin scintillator read out by SiPMs in a way to reconstruct the position and shape of the scintillation site, hence the beam spot.
    
    The easiest geometry, presented in this paper, consists of a thin scintillator coupled with a 2D SiPM matrix. However, with this setup the SiPMs are exposed directly to the muons and might go through a radiation damage process\cite{Garutti2018Radiation, Klanner2018Characterisation}. 
    In particular, muons are expected to cause lattice defects (displacements), increasing the number of generation-recombination centres, which eventually yields an increase of the dark count rate, hence an increase of the dark current. Thermal annealing (\emph{e.g.} 24 hours at 125 $^\circ$C) allows partial recombination of lattice defects, with an expected reduction of the measurable damage in terms of dark current and dark count rate.
    In addition, muons are not expected to modify the junction effective doping during their interaction with the nuclei, therefore, no variation in the breakdown voltage is expected. 
    Some preliminary evidence of radiation damage of SiPMs exposed to low-momentum muon beams is presented in this work, even though extensive testing is still ongoing. 
    
\section{Test beam with muons at ISIS RIKEN-RAL Port4}
    \begin{figure}
        \centering
        \includegraphics[width=\linewidth]{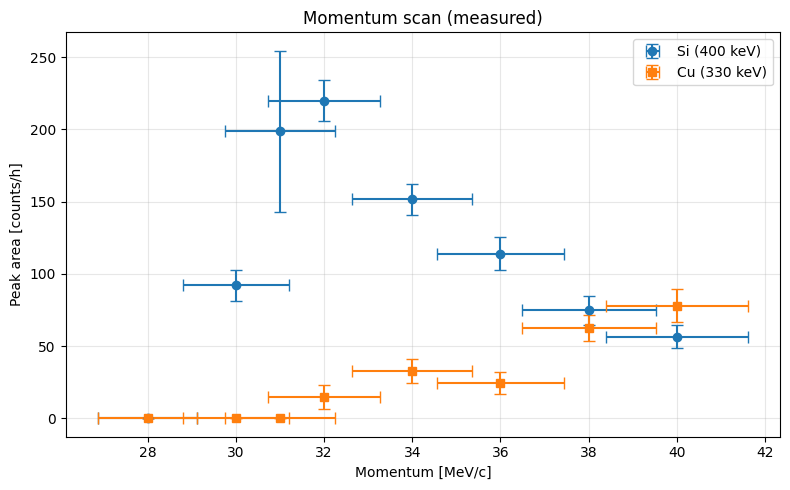}
        \caption{Count rate for the main silicon and copper muonic X-rays as a function of beam momentum when irradiating the SiPM matrix with $\upmu^-$. The maximum energy deposit in the SiPM occurs at 31-32 MeV/c. The uncertainty bars are statistical, the point at 31 MeV/c was acquired with $\sim 30$\% the statistics with respect to the other points. The uncertainties on the momentum correspond to the 4\% momentum bite.}
        \label{fig:mom-scan}
    \end{figure}

    \begin{table}[]
        \centering
        \caption{List of the SiPM models characterised in this study. The size of all SiPMs is $3 \times 3$ mm.}
        \label{tab:sipm}
        \begin{tabular}{c|c|c}
             \toprule
             \textbf{Manufacturer} & \textbf{Model} & \textbf{Cell size ($\upmu$m)}\\
             \midrule
             Hamamatsu & S14160-3050PS & 50 \\
             Hamamatsu & S13360-3025PE & 25 \\
             Hamamatsu & S13360-3050PE & 50 \\
             Hamamatsu & S13360-3075PE & 75 \\
             AdvanSiD & NUV & 40 \\
             Onsemi (SensL) & C-series 3 mm& 35 \\
             Broadcom & AFBR-S4N44C013 & 30 \\
             KETEK & PM3325-WB-D0 & 25 \\
             \bottomrule
        \end{tabular}
    \end{table}
    
    \begin{figure*}
        \centering
        \includegraphics[height=0.32\linewidth]{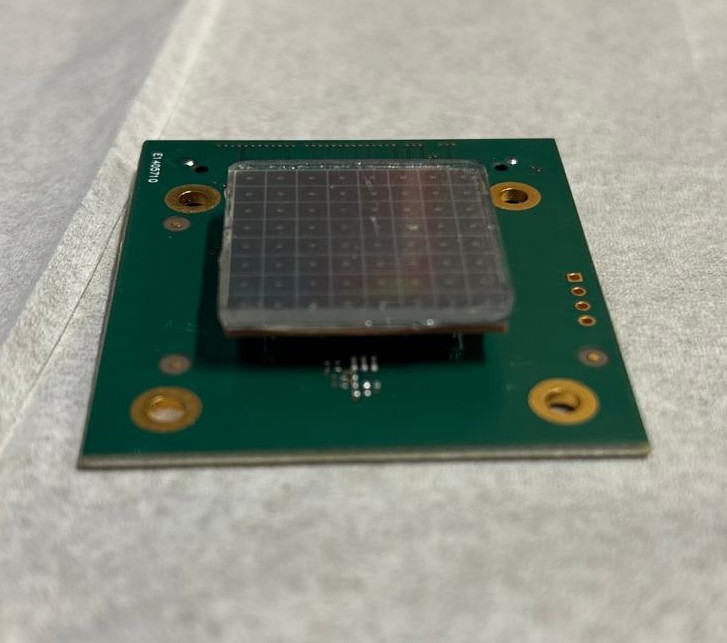}
        \includegraphics[height=0.32\linewidth]{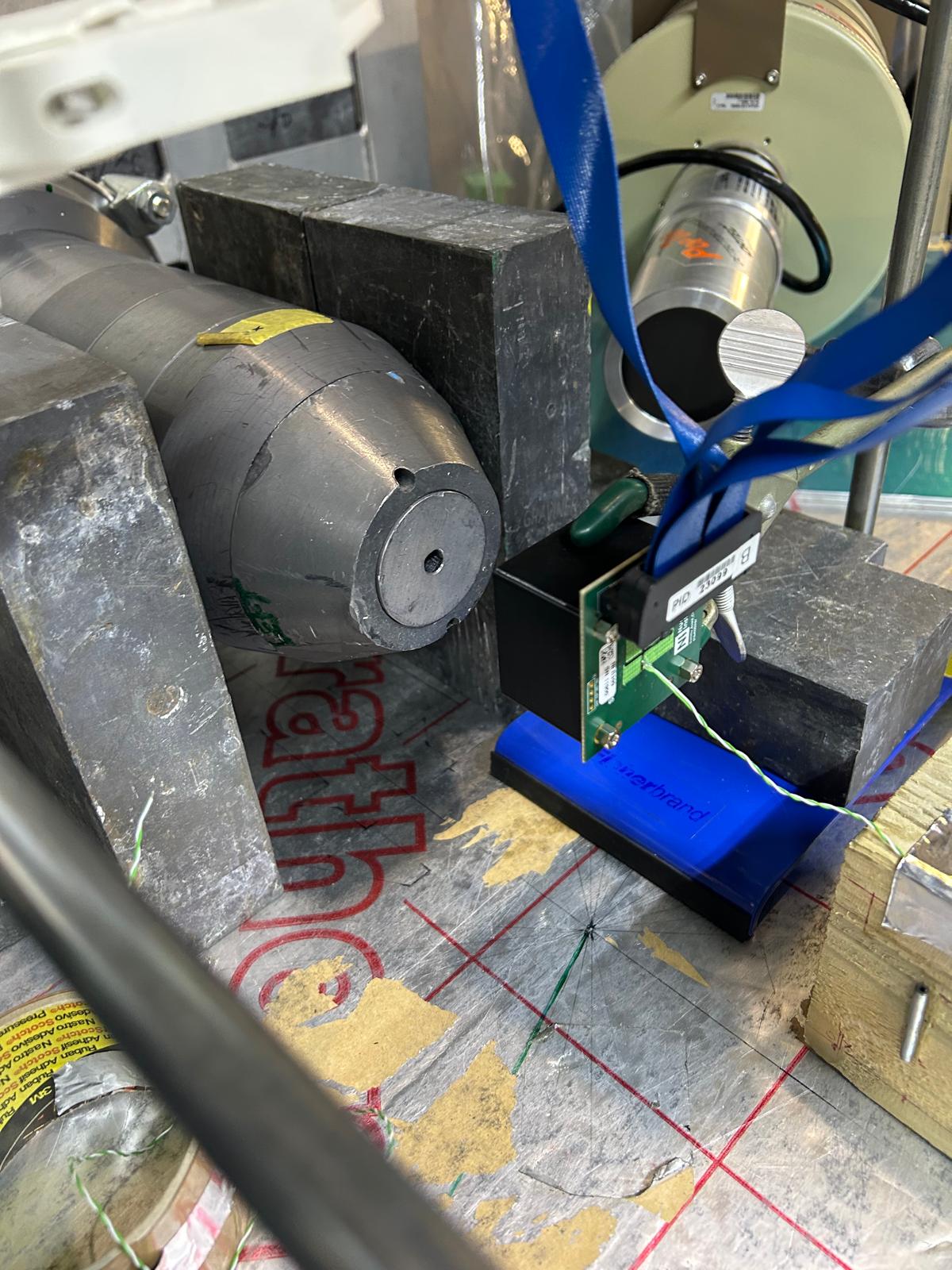}
        \includegraphics[height=0.32\linewidth]{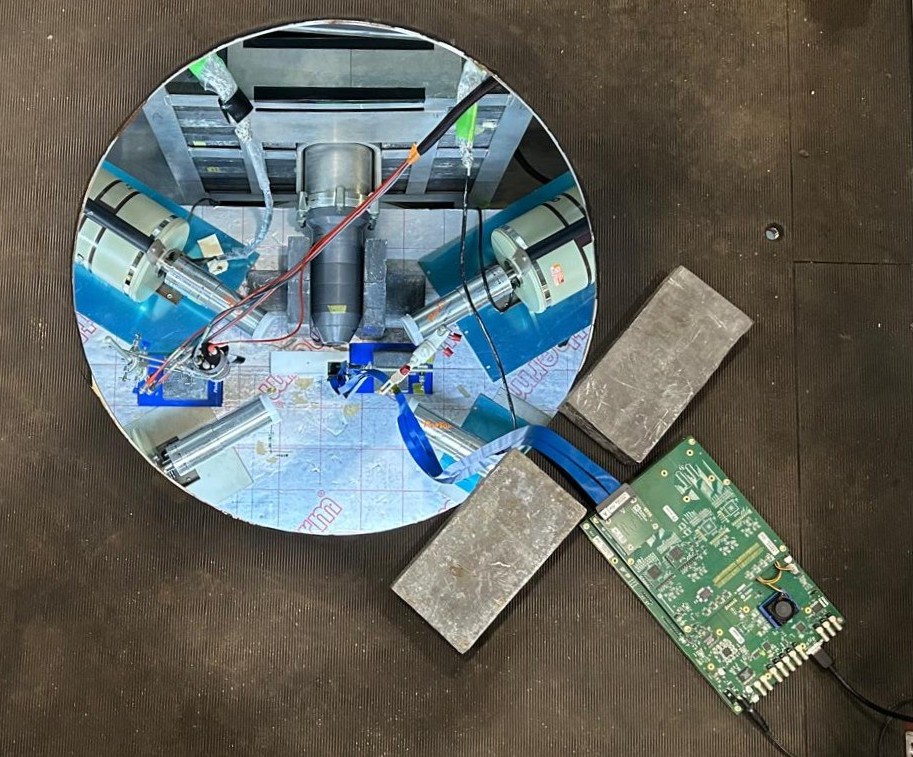}
        \caption{Experimental setup for the test beam at RIKEN-RAL Port4: SiPM matrix coupled with a 2 mm-thick plastic scintillator (\textit{left}), detector prototype mounted in front of the beam collimator (\textit{centre}), and setup seen from above with four HPGe detectors and the CAEN DT5550W-A55PET2 board (\textit{right}).}
        \label{fig:setup}
    \end{figure*}

      \begin{figure}
        \centering
        \includegraphics[width=0.8\linewidth]{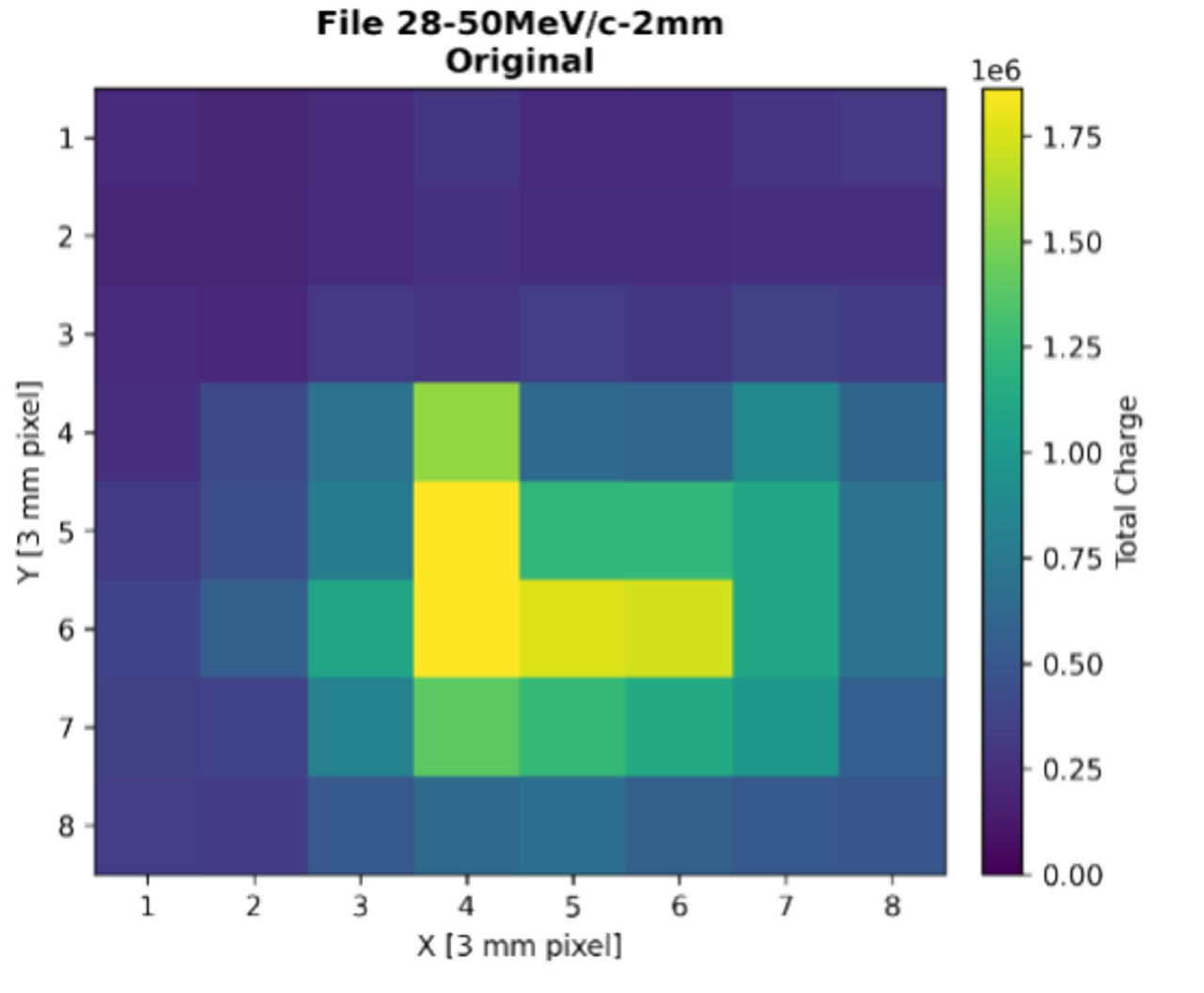}
        \caption{Preliminary 2D beam spot determination (beam of $\upmu^-$ at 50 MeV/c, with an 8 mm circular collimator) obtained by requiring events to have trigger on a single SiPM. Each pixel (bin) corresponds to a SiPM, $3 \times 3$ mm$^2$. The beam is originally elliptical and focused upstream of the focus, it is indeed visible that the beam spot is wider on the $x$ axis.}
        \label{fig:pixel}
    \end{figure}

    \begin{figure}
        \centering
        \includegraphics[width=\linewidth]{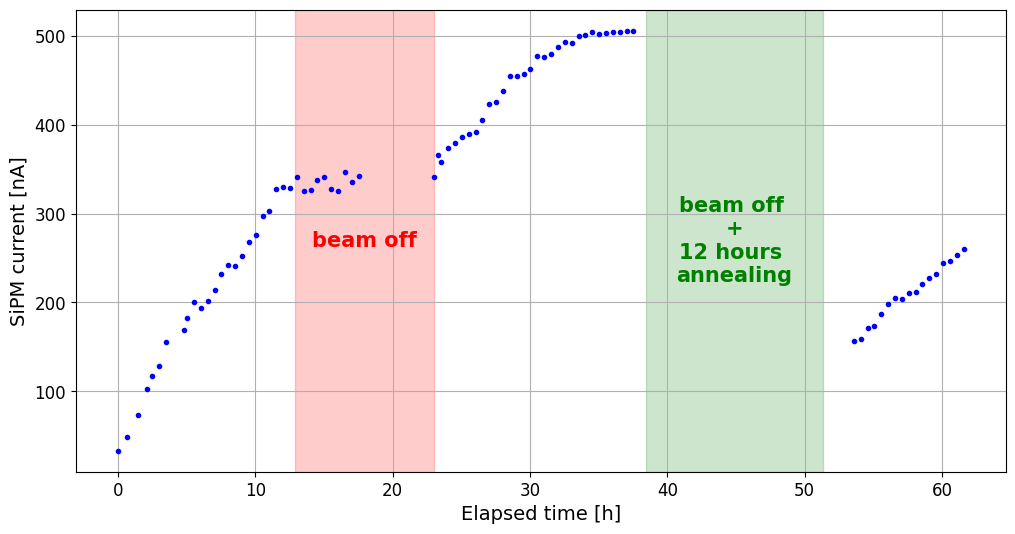}
        \caption{Live dark current values for a Hamamatsu S14160-3050HS exposed for 3 nights to maximum intensity surface muons. After $\sim$12 hours without beam no recovery is observed, whereas after $\sim$12 hours annealing at 125 $^\circ$C partial recovery is observed.}
        \label{fig:irrad-live}
    \end{figure}
    
    Three days of beamtime at RIKEN-RAL Port4 were granted to carry out the studies presented in this work. 

    A set of Hamamatsu S14161-3050HS-08 matrices, each consisting of 64 SiPMs ($8 \times 8$) with $3 \times 3$ mm$^2$ active area and 50 $\upmu$m cell, was used for the test. Each SiPM mounted a SAMTEC SS4-40-3.00-L-DK-TR connector to plug it into the prototype PCB, which was connected to the CAEN DT5550W-A55PET2 voltage supply and readout system through a cable. This board can operate up to 64 channels providing the bias and front-end readout through two ASIC PETIROC2A. In particular, each channel operates directly on the analog data through two simultaneous operational branches: a fast shaper for triggering (intrinsic resolution $\sim 37$ ps) and a slow shaper which feeds its output to the FPGA for energy and time estimation. These values are then stored. Figure \ref{fig:setup} shows the setup mounted on the beamline. 
    
    The setup has initially been scanned using decay $\upmu^-$ of various momenta to study the energy deposit in the layers of the prototype. This study can be done with Muonic atom X-ray Emission Spectroscopy ($\upmu$-XES)\cite{rossini2024hodo, cataldo, ninomiya, tandem, meteorites}, which consists of a bulk elemental characterisation of a sample by measuring the X and $\gamma$ rays emitted after $\upmu^-$ irradiation. Negative muons stopped in a target form muonic atoms, emitting characteristic X-rays, and can then be captured by the nucleus emitting characteristic $\gamma$ rays, all used for elemental identification. As the muon energy loss is described by the Bethe-Bloch formula, it loses energy with a Bragg peak. For this reason, scanning a sample as a function of momentum lets experimenters understand its composition at the studied layer. In our case, scanning the prototype lets us find the momentum at which the energy deposit is maximised in the SiPM. 

    At RIKEN-RAL Port4, the scan was done using the MuX facility germanium detectors. A result of the momentum scan between 28 MeV/c and 40 MeV/c is shown in Figure \ref{fig:mom-scan}. The prototype detector was then tested with a beam of negative muons at 50 MeV/c and a circular collimator with radius of 8 mm, resulting in the preliminary beam profile shown in Figure \ref{fig:pixel}.

    In addition, the SiPM models listed in Table \ref{tab:sipm} were tested for radiation damage effects, by exposing them to the same fluence and the same kind of muons. The maximum deposit in Si inside the plastic case (see Figure \ref{fig:mom-scan}) is at 30-31 MeV/c, almost twice the deposit at 29 MeV/c. However, surface muon intensity at 29 MeV/c is two orders of magnitude greater than that of 30-31 MeV/c decay muons. It has therefore been decided to expose SiPMs to surface muons. The beam profile can be assumed uniform over the measured beam spot ($\sim 36 \times 24$ mm$^2$) with intensity $10^6 \ \upmu^+$/s, and each SiPM was exposed to the beam for 13 hours at full intensity. For this reason, the estimated fluence on each SiPM (9 mm$^2$) is roughly $\sim 5.4 \cdot 10^9 \ \upmu^+$/cm$^2$, corresponding to $\sim 5 \cdot 10^8$ muons deposited on each SiPM. In addition, as many muons are expected to stop in the SiPM active area, possibly in the depletion region, a contribution to the dose is also given by positrons emitted during the muon decay, causing further damage. 
    
    SiPM irradiation was carried out at night, while the detector prototype testing was carried out during the day. 

    Each SiPM was characterised in terms of $I$-$V$ characteristic curve and waveform acquisition before and after irradiation. Each acquisition was carried out in a climatic chamber at 25 $^\circ$C, the results are discussed in the next Section along with the thermal annealing.
    
    To monitor the ongoing damage, another SiPM Hamamatsu S14160-3050PS (50 $\upmu$m cell) was exposed to the beam at operating voltage, monitoring the dark current to qualitatively check the generation of radiation-induced defects in real time. The SiPM was thoroughly wrapped and checked to exclude external light contamination. For the first two nights, the dark current kept increasing, even after the daytime without beam. During the second day, the SiPM was put in a muffle furnace at 125 $^\circ$C for 12 hours to check the effect of thermal annealing. Indeed, during the third night, the SiPM had partially recovered and started from a lower dark current. The plot of dark current over time is shown in Figure \ref{fig:irrad-live}. No quantitative information can be extracted from this plot as the SiPM was exposed without active temperature-bias correction, and the absolute values could be affected by temperature and gain drift. 


\section{Test of SiPM annealing}

    \begin{figure}
        \centering
        \includegraphics[width=\linewidth]{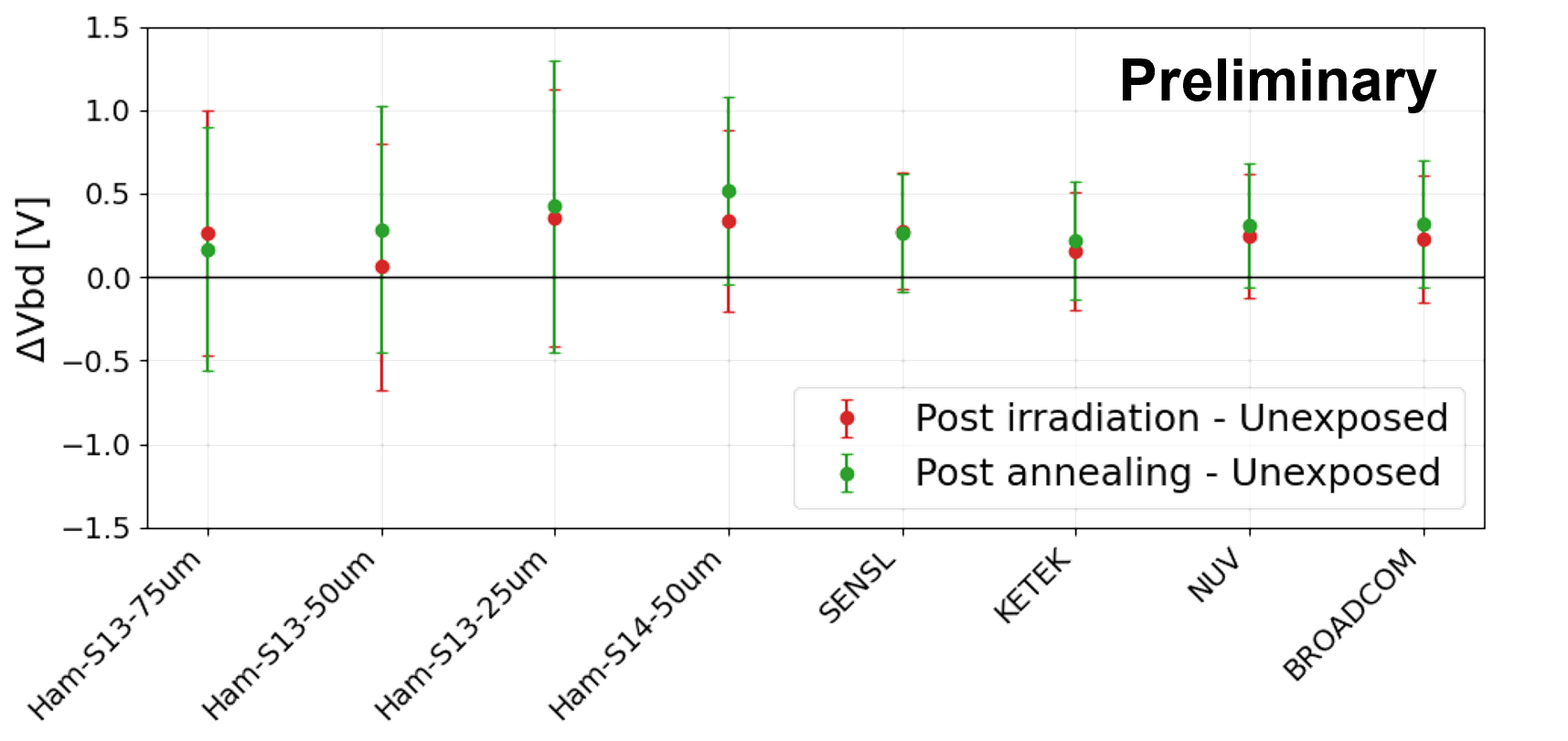}
        \caption{Comparison of the initial, after irradiation, and after annealing values of the breakdown voltage. It is currently under investigation whether the slightly higher value observed systematically for all SiPMs is due to residual neutrons in the beam\cite{monza:2026}.}
        \label{fig:Vbr}
    \end{figure}

    After the irradiation, all eight irradiated SiPMs have undergone an established procedure\cite{Huang2025:annealing, Dolenec2025:annealing, Cordelli2018:annealing} of thermal annealing consisting of heating the SiPMs at 125 $^\circ$C for 24 hours, which is a good balance to have fast annealing without breaking the devices. The SiPMs have then been tested again in terms of $I$-$V$ characteristic curve and waveform acquisition.

    In this work, only the results on the $I$-$V$ characteristic curves are provided, as work on the dark count rate estimation is still ongoing. The breakdown voltage is estimated from the $I$-$V$ curve by fitting $\sqrt{I}(V)$ with two linear regressions in the leakage current region and in the first part of the dark current region, after the breakdown voltage, which is defined from the intersection of the two lines\cite{Klanner2018Characterisation, hodomaxi}. 

    Figure \ref{fig:Vbr} shows a comparison for all SiPMs of the breakdown voltage before irradiation, after irradiation and after annealing. No clear change is observed, within the calculated tolerances, in the three phases. This confirms that no pn junction effective doping modification is expected to be observed for positive muon irradiation. On the other hand, nuclear negative muon capture ($\mu^- \ p \rightarrow n \ \nu_\mu $) could in principle transmute Si or dopant nuclei and modify the pn junction.
    
    However, post-irradiation and post-annealing data show a small but systematic increase in breakdown voltage with respect to unexposed SiPMs, suggesting that either some effect is caused by muons or secondary particles, or some residual neutrons are also transported in the final part of the beamline. At RIKEN-RAL Port4, the CHNet-MAXI experiment\cite{hodomaxi} is currently testing CLLB scintillators, optimised to discriminate neutrons\cite{monza:2026}, which could help in understanding the cause of this effect, if observed systematically. 

    \begin{figure}
        \centering
        \includegraphics[width=\linewidth]{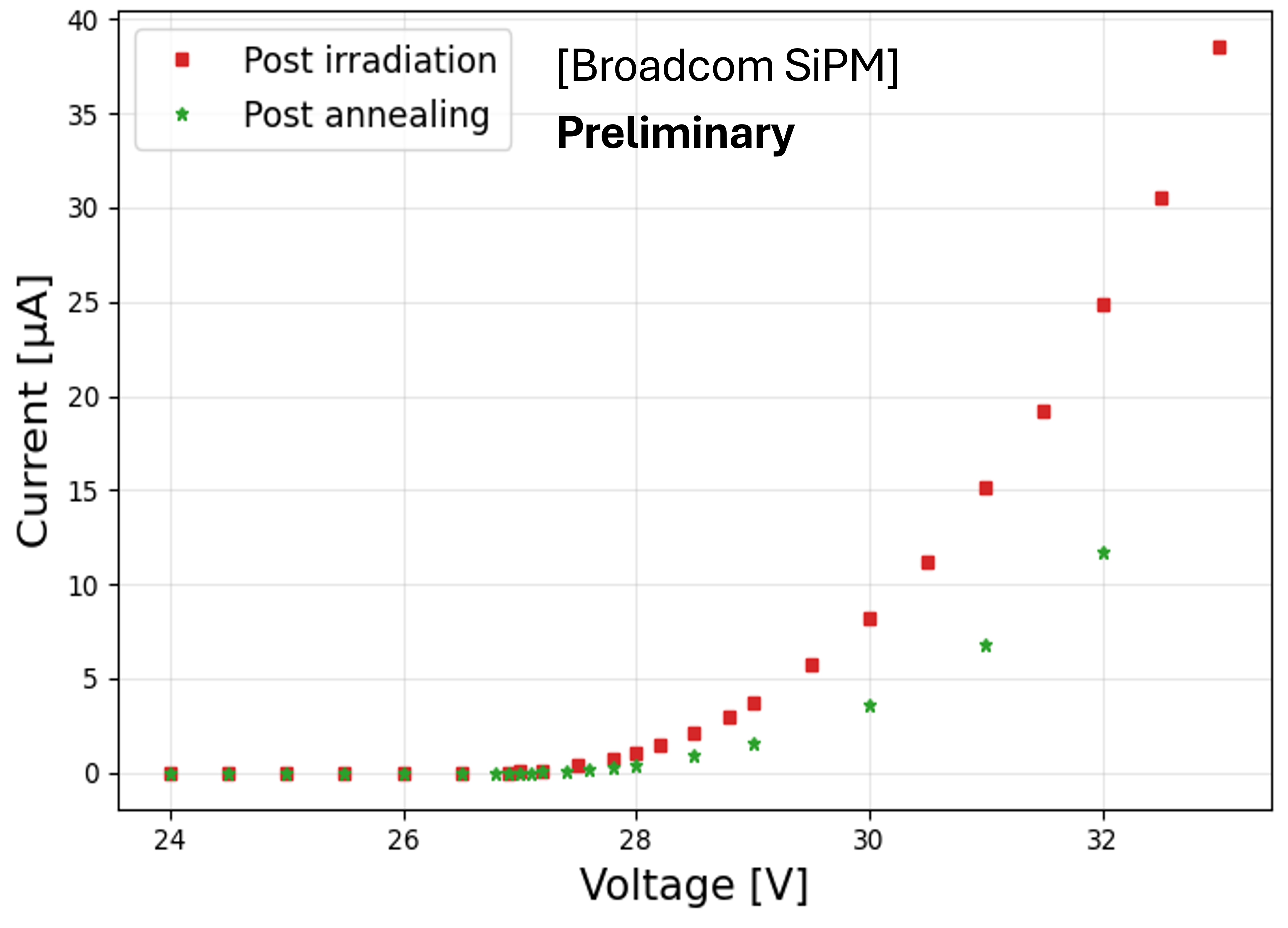}
        \caption{Example of $I$-$V$ characteristic curve after irradiation (red squares) and after annealing (green stars), showing compatible breakdown voltage and a reduced dark current after the annealing with respect to after irradiation.}
        \label{fig:bro}
    \end{figure}

\section{Conclusion}

    In order to monitor low-momentum ($\lesssim$ 40 MeV/c) muon beams with a dense detector, the research and development on thin scintillating detectors is ongoing. 

    The easiest geometry consists of a thin scintillator read out with a 2D SiPM matrix, for which the feasibility was tested. The main downside of this geometry is that the SiPMs are exposed directly to the muon beam. 

    For this reason, various SiPM models were exposed to the RIKEN-RAL surface muon beam at ISIS, indicating that the damage is mostly recoverable through thermal annealing, which will be further investigated by our studies on the dark count rate.

\section*{Acknowledgements}

The work was supported by Fondazione Angelo Della Riccia (International Exchange Grant 2024-2025 awarded to R.R.) and the Royal Society (International Exchange Grant no. IES-R2-244430 awarded to J.S.L. and A.M.), both of which are gratefully acknowledged for their contributions.

The muon beam was provided by the Science and Technology Facilities Council (STFC) through access to the ISIS Neutron and Muon Source under the CNR-STFC Agreement 2021–2027. Beamtime was allocated under proposal RB2520455\footnote{\url{https://doi.org/10.5286/ISIS.E.RB2520455}}.

Roberto Bertoni (INFN Milano-Bicocca) and Sun Wen (University of Nanjing) are gratefully acknowledged for the technical support. 



\bibliographystyle{elsarticle-num}
\bibliography{biblio}
\end{document}